\documentclass{article}
\usepackage{graphics}
\newtheorem{thmc}{Theorem}
\newtheorem{prop1}[thmc]{Proposition}
\newtheorem{prop2}[thmc]{Proposition}
\newtheorem{prop0}[thmc]{Proposition}
\newtheorem{lemma1}[thmc]{Lemma}
\newtheorem{lemma2}[thmc]{Lemma}

\newtheorem{lemma4}[thmc]{Lemma}

\begin{document}

\def\tr{ {\rm{Tr \,}}}
\def\diag{ {\rm{diag \,}}}

\def\supp{ {\rm{supp \,}}}
\def\dim{ {\rm{dim \,}}}
\def\oti{{\otimes}}
\def\bra#1{{\langle #1 |  }}
\def\lb{ \left[ }
\def\rb{ \right]  }
\def\tilde{\widetilde}
\def\bar{\overline}
\def\*{\star}

\def\({\left(}		 		 \def\BL{\Bigr(}
\def\){\right)}		 		 \def\BR{\Bigr)}
		 \def\BBL{\lb}
		 \def\BBR{\rb}
%

\def\bb{{\bar{b} }}
\def\ab{{\bar{a} }}
\def\zb{{\bar{z} }}
\def\zbar{{\bar{z} }}
\def\frac#1#2{{#1 \over #2}}
\def\inv#1{{1 \over #1}}
\def\half{{1 \over 2}}
\def\d{\partial}
\def\der#1{{\partial \over \partial #1}}
\def\dd#1#2{{\partial #1 \over \partial #2}}
\def\vev#1{\langle #1 \rangle}
\def\ket#1{ | #1 \rangle}
\def\proj#1{ | #1 \rangle \langle #1 |}
\def\rvac{\hbox{$\vert 0\rangle$}}
\def\lvac{\hbox{$\langle 0 \vert $}}
\def\2pi{\hbox{$2\pi i$}}
\def\e#1{{\rm e}^{^{\textstyle #1}}}
\def\grad#1{\,\nabla\!_{{#1}}\,}
\def\dsl{\raise.15ex\hbox{/}\kern-.57em\partial}
\def\Dsl{\,\raise.15ex\hbox{/}\mkern-.13.5mu D}
\def\b#1{\mathbf{#1}}
%
%
\def\th{\theta}		 		 \def\Th{\Theta}
\def\ga{\gamma}		 		 \def\Ga{\Gamma}
\def\be{\beta}
\def\al{\alpha}
\def\ep{\epsilon}
\def\vep{\varepsilon}
\def\la{\lambda}		 \def\La{\Lambda}
\def\de{\delta}		 		 \def\De{\Delta}
\def\om{\omega}		 		 \def\Om{\Omega}
\def\sig{\sigma}		 \def\Sig{\Sigma}
\def\vphi{\varphi}
%
%
\def\CA{{\cal A}}		 \def\CB{{\cal B}}		 
\def\CC{{\cal C}}
\def\CD{{\cal D}}		 \def\CE{{\cal E}}		 
\def\CF{{\cal F}}
\def\CG{{\cal G}}		 \def\CH{{\cal H}}		 
\def\CI{{\cal J}}
\def\CJ{{\cal J}}		 \def\CK{{\cal K}}		 
\def\CL{{\cal L}}

\def\CM{{\cal M}}		 \def\CN{{\cal N}}		 
\def\CO{{\cal O}}
\def\CP{{\cal P}}		 \def\CQ{{\cal Q}}		 
\def\CR{{\cal R}}
\def\CS{{\cal S}}		 \def\CT{{\cal T}}		 
\def\CU{{\cal U}}
\def\CV{{\cal V}}		 \def\CW{{\cal W}}		 
\def\CX{{\cal X}}
\def\CY{{\cal Y}}		 \def\CZ{{\cal Z}}
\newcommand{\qed}{\rule{7pt}{7pt}}
\def\E{{\mathbf{E} }}
\def\1{{\mathbf{1} }}
\def\rvac{\hbox{$\vert 0\rangle$}}
\def\lvac{\hbox{$\langle 0 \vert $}}
\def\comm#1#2{ \BBL\ #1\ ,\ #2 \BBR }
\def\2pi{\hbox{$2\pi i$}}
\def\e#1{{\rm e}^{^{\textstyle #1}}}
\def\grad#1{\,\nabla\!_{{#1}}\,}
\def\dsl{\raise.15ex\hbox{/}\kern-.57em\partial}
\def\Dsl{\,\raise.15ex\hbox{/}\mkern-.13.5mu D}
\def\beq{\begin {equation}}
\def\eeq{\end {equation}}
\def\to{\rightarrow}

\title{The capacity of a quantum channel for simultaneous transmission 
of classical and quantum information}

\author{
I. Devetak\\
\it{IBM T.J. Watson Research Center, Yorktown Heights, NY 10598}\\ 
\\
P. W. Shor\\
\it{Department of Mathematics, MIT, Cambridge, MA 02139} \\
} 

\date{\today} 
\maketitle

\begin{abstract}
An expression is derived characterizing the set of admissible rate pairs 
for simultaneous transmission of classical and quantum 
information over a given quantum channel, generalizing both the classical 
and quantum capacities of the channel. Although our formula involves
regularization, i.e. taking a limit over many copies of the channel,
it reduces to a single-letter expression in the case of generalized dephasing channels.
Analogous formulas are conjectured for the simultaneous public-private capacity 
of a quantum channel and for the simultaneously 1-way 
distillable common randomness and entanglement of a bipartite quantum state.
   
\end{abstract}

\section{Introduction}

In the paper that marked the beginning of information theory 
\cite{shannon}, 
C. E. Shannon introduced the notion of a (classical) channel $W$, 
a stochastic map modeling
the effect of noise experienced by a classical message on its
way from sender to remote receiver. There he defined and computed
the key property of the channel $W$: its \emph{capacity} $C(W)$ to
convey classical information, expressed in bits per channel use. 
Many decades later, in the context of quantum information theory,
the notion of a \emph{quantum channel} $\CN$, a cptp (completely positive 
trace
preserving) map, was introduced as the most general bipartite dynamic 
resource consistent with quantum mechanics. 
There are now two basic capacities one may define for $\CN$: classical 
$C(\CN)$ and quantum $Q(\CN)$. Intuitively, these correspond to
the maximum number of bits (respectively qubits) per use of $\CN$ that can
be faithfully transmitted over the channel.
The classical capacity theorem was independently
proved by Holevo \cite{h}, and Schumacher and Westmoreland \cite{sw}. The
quantum capacity theorem was originally stated by Lloyd \cite{lloyd},
although it was only recently generally realized that his proof could be
made rigorous \cite{horodecki-lloyd}.  It has also been
proved by Shor \cite{shor}
and subsequently, via the private classical capacity, by Devetak \cite{key}. 
In the present paper we unify the two capacities by investigating the 
capacity of $\CN$ for \emph{simultaneously} transmitting classical and quantum
information, given in the form of a trade-off curve. 

Let the sender Alice and receiver Bob be connected via a quantum channel
$\CN: \CH_{A'} \rightarrow \CH_{B}$, where
$\CH_{A'}$ denotes the Hilbert space of Alice's
input system $A'$ and $\CH_{A'}$ that of Bob's output
system $B$.
We shall define three distinct information processing scenarios which
will turn out to be equivalent.

\paragraph{Scenario Ia (subspace transmission)} Alice's task is to convey 
to Bob, 
in some large number $n$ uses of the channel, 
one of $\mu$ equiprobable classical messages
with low error probability and simultaneously an arbitrarily chosen
quantum state from some Hilbert space $\CH$ of dimension
$\kappa$ with high fidelity.
More precisely, we define a (classical, quantum)  \emph{channel code}  
to consist of:

$\bullet$ An ordered set $(\CE_m)_m \in [\mu]$, 
$[\mu] = \{ 1,2, \dots \mu \}$, of cptp maps
$\CE_m: \CH_{A''} \rightarrow \CH_{A'}^{\otimes n}$. 
Such an ordered set the most general function
with two inputs, classical and quantum, and one quantum output.

$\bullet$ A decoding quantum instrument \cite{instrument} 
$\mathbf{D} = (\CD_m)_{m \in 
[\mu]}$,
an ordered set of cp (completely positive) maps 
$\CD_m: \CH_{B}^{\otimes n} \rightarrow \CH_{B'}$, 
the sum of which $\CD = \sum_{m \in [\mu]} \CD_m$ is trace preserving.
The probability of outcome $m$ for input $\rho$ is $\tr \CD_m(\rho)$,
while the effective quantum map is $\CD$. 
The instrument has one quantum input and two outputs, classical and 
quantum.
It is a natural generalization of a POVM (positive operator valued measure), 
which cares only about the 
classical output, and quantum cptp map, which only has a 
quantum output.  

\vspace{2mm}

Alice's classical message is represented by a random variable $M$
uniformly distributed on the set $[\mu]$. Conditional
on $M$ taking on a particular value $m$,
Alice encodes the quantum state of $A''$ with $\CE_m$ and sends
it through $n$ copies of the channel $\CN$. 
Bob performs the instrument $\mathbf{D}$ on the channel output,
resulting in the classical outcome random variable $M'$ and
a quantum output system $B'$. Note that $\CH_{A''} = \CH_{B'} = \CH$.
We call the ordered pair $((\CE_m)_m, \mathbf{D})$ an $(n,\epsilon)$ code if
\begin{eqnarray*}
\lefteqn{  1. \, \Pr \{M'  \neq m| M = m \} \leq  \epsilon,  \,\,\,\,\,\,  \forall m,}
\,\,\,\,\,\,\,\,\,\,\,\,\,\,\,\,\,\,\,\,\,\,\,\,\,\,\,\,\,\,\,\,\,\,\,
\,\,\,\,\,\,\,\,\,\,\,\,\,\,\,\,\,\,\,\,\,\,\,\,\,\,\,\,\,\,\,\,\,\,\,
\,\,\,\,\,\,\,\,\,\,\,\,\,\,\,\,\,\,\,\,\,\,\,\,\,\,\,\,\,\,\,\,\,\,\,
\,\,\,\,\,\,\,\,\,\,\,\,\,\,\,\,\,\,\,\,\,\,\,\,\,\,\,\,\,\,\,\,\,\,\,
\,\,\,\,\,\,\,\,\,\,\,\,\,\,\,\,\,\,\,\,\,\,\,\,\,\,\,\,\,\,\,\,\,\,\,
\,\,\,\,\,\,\,\,\,\,\,\,\,\,\,\,\,\,\,\,\,\,\,\,\,\,\,\,\,\,\,\,\,\,\,
 \label{jen} \\ 
\lefteqn{ 2.  \, \min_{\ket{\varphi} \in \CH} F({\varphi},
 (\CD \circ \CN^{\otimes n} \circ \CE_m) ({\varphi})) 
\geq  1 - \epsilon, \,\,\,\,\,\, \forall m,} 
\,\,\,\,\,\,\,\,\,\,\,\,\,\,\,\,\,\,\,\,\,\,\,\,\,\,\,\,\,\,\,\,\,\,\,
\,\,\,\,\,\,\,\,\,\,\,\,\,\,\,\,\,\,\,\,\,\,\,\,\,\,\,\,\,\,\,\,\,\,\,
\,\,\,\,\,\,\,\,\,\,\,\,\,\,\,\,\,\,\,\,\,\,\,\,\,\,\,\,\,\,\,\,\,\,\,
\,\,\,\,\,\,\,\,\,\,\,\,\,\,\,\,\,\,\,\,\,\,\,\,\,\,\,\,\,\,\,\,\,\,\,
\,\,\,\,\,\,\,\,\,\,\,\,\,\,\,\,\,\,\,\,\,\,\,\,\,\,\,\,\,\,\,\,\,\,\,
\,\,\,\,\,\,\,\,\,\,\,\,\,\,\,\,\,\,\,\,\,\,\,\,\,\,\,\,\,\,\,\,\,\,\,
\label{uvjet}
\end{eqnarray*}
where the \emph{fidelity} is defined by 
$F(\rho, \sigma) = \|\sqrt{\rho} \sqrt{\sigma} \|_1^2$.
Condition 1 above means that each message should be 
correctly decoded by Bob with high probability. 
Condition 2  corresponds to the 
\emph{subspace transmission} criterion of \cite{bkn}:
each pure input state $\ket{\phi}$ supported on $\CH$ 
should be almost perfectly transmitted to Bob.
The (classical, quantum) \emph{rate pair} of the code is $(r, R)$,
with $r = \frac{1}{n} \log \mu$ and $R = \frac{1}{n} \log \kappa$.
They represent the number of bits and qubits, respectively, per use of the
channel that can be faithfully transmitted simultaneously. 
A rate pair $(r, R)$ is called \emph{achievable} if for all $\epsilon, 
\delta > 0$
and all sufficiently large $n$ there exists an $(n, \epsilon)$ code with rate
pair $(r - \delta, R - \delta)$.
The simultaneous (classical, quantum) scenario Ia capacity region 
of the channel $S_{\rm{Ia}}(\CN)$ is the set of all achievable positive rate pairs.

\paragraph{Scenario Ib (entanglement transmission)} 

This scenario is very similar to the first one, but instead
of transmitting an arbitrary pure state of $A''$, Alice is
required to preserve entanglement \cite{bkn} between $A''$
and some reference system $A$ she has no access to. 
Here condition 2 is replaced by 
\begin{eqnarray*}
\lefteqn{ 2'.  \, F({\Phi}, \Omega_m) \geq 1 - \epsilon,
\,\,\,\,\,\, \forall m,} 
\,\,\,\,\,\,\,\,\,\,\,\,\,\,\,\,\,\,\,\,\,\,\,\,\,\,\,\,\,\,\,\,\,\,\,
\,\,\,\,\,\,\,\,\,\,\,\,\,\,\,\,\,\,\,\,\,\,\,\,\,\,\,\,\,\,\,\,\,\,\,
\,\,\,\,\,\,\,\,\,\,\,\,\,\,\,\,\,\,\,\,\,\,\,\,\,\,\,\,\,\,\,\,\,\,\,
\,\,\,\,\,\,\,\,\,\,\,\,\,\,\,\,\,\,\,\,\,\,\,\,\,\,\,\,\,\,\,\,\,\,\,
\,\,\,\,\,\,\,\,\,\,\,\,\,\,\,\,\,\,\,\,\,\,\,\,\,\,\,\,\,\,\,\,\,\,\,
\,\,\,\,\,\,\,\,\,\,\,\,\,\,\,\,\,\,\,\,\,\,\,\,\,\,\,\,\,\,\,\,\,\,\,
\end{eqnarray*}
where
\beq
\Omega^{A B'}_m = [\1^{A} \otimes (\CD \circ \CN^{\otimes 
n} \circ \CE_m)]({\Phi}^{A A''}),
\eeq
and 
\beq
\ket{\Phi}  =  \sqrt{\frac{1}{\kappa}} \sum_{k= 1}^{\kappa} \ket{k} \otimes
\ket{k}
\eeq
is the standard maximally entangled state on $\CH \otimes \CH$. 
We denote the corresponding capacity 
region by $S_{\rm{Ib}}(\CN)$. 

\paragraph{Scenario II (entanglement generation)} 
In this scenario, simultaneously with transmitting
classical information, Alice wishes to \emph{generate entanglement}
\cite{key} shared
with Bob rather than preserving it as in scenario Ib.
Alice prepares, without loss of generality, a pure 
bipartite state $\ket{\Upsilon_m}^{A A'^n}$ 
in her lab ($\CH_{A'^n}:= \CH_{A'}^{\otimes n}$),
depending on the classical information $m$, and sends
it through the channel. Bob decodes as above, yielding the output
state
\beq
\Omega^{A B'}_m = [\1^{A} \otimes (\CD \circ \CN^{\otimes 
n})]({\Upsilon_m}^{A A'^n}),
\eeq
shared by Alice and Bob. Everything else is defined as in scenario Ib.
The corresponding capacity region is denoted by $S_{\rm{II}}(\CN)$.
\vspace{2mm}

In the next section we state our main result, a unique expression
for the capacity regions defined above, investigate its properties
and relate it to previous work. The proof of our main theorem is relegated to
section 3. Some remarks on related problems are collected in section 4.
We conclude in section 5 with suggestions for future research.

\section{Main result}

Recall the notion of an \emph{ensemble}
of quantum states $E = \{p_x, \ket{\phi_x}^{A A'}\}$:
the quantum system $A A'$ is in the 
state $\ket{\phi_x}^{A A'}$ with probability $p_x$.
The ensemble $E$ is equivalently represented by a
\emph{classical-quantum} system \cite{CR} $X A A'$
in the state
$$
\sum_x p_x \ket{x} \bra{x}^X \otimes \ket{\phi_x} \bra{\phi_x}^{A A'}.
$$
$X$ plays the dual role of 
an auxiliary quantum system in the
state $\sum_x p_x \ket{x} \bra{x}$ and of 
a random variable with distribution $p$.
Sending the $A'$ system through the channel $\CN$ gives
rise to a classical-quantum system $X A B$ in some state $\sigma^{X A B}$:
\beq
\sigma^{X A B} = \sum_x p_x \ket{x} \bra{x}^X \otimes 
[(\1 \otimes \CN)(\ket{\phi_x} \bra{\phi_x})]^{A B}.
\label{szigma}
\eeq
For such a state we say that it ``arises from'' the channel $\CN$.
For a multi-party state such as $\sigma^{XAB}$ the
reduced density operator $\sigma^{A}$ is defined by 
$\tr_{\!XB} \sigma^{XAB}$. Conversely, we 
call $\sigma^{XAB}$ an \emph{extension} of $\sigma^A$.
A pure extension is conventionally called a 
\emph{purification}.
Define the von Neumann entropy of a quantum state $\rho$
by $H(\rho) = - \tr( \rho \log \rho)$. 
We write $H(A)_\sigma = H(\sigma^A)$,
omitting the subscript when the reference state is clear from the context.
The Shannon entropy $-\sum_x p_x \log p_x$ 
of the random variable $X$ is equal to the von Neumann entropy $H(X)$
of the system $X$. 
Define the 
conditional entropy 
$$
H(A|B) = H(B) - H(AB),
$$
(quantum) mutual information 
$$
I(A;B) = H(A) + H(B) - H(AB),
$$
and conditional mutual information
$$
I(A;B|X) = I(A;BX) - I(A;X).
$$
The \emph{coherent information} $I(A\, \rangle B)$
is defined as $- H(A|B)$. 
Whenever the state  $\rho^{AB}$ comes about by sending 
some pure state $\ket{\phi}^{A A'}$ through the channel $\CN$,
we may use the alternative notation \cite{sn}
$I_c(\phi^{A'}, \CN) := I(A\,\rangle  B)_{\rho}$, since 
this quantity is independent of the particular purification
$\ket{\phi}^{A A'}$  of $\phi^{A'}$.
In what follows all information theoretical quantities
 will refer to the state $\sigma^{X A B}$,
unless stated otherwise.

Our main result is the following theorem.

\vspace{1mm}

\begin{thmc} 
The simultaneous capacity regions of $\CN$ for the various scenarios 
Ia, Ib and II are all equal and  given by
\beq
S(\CN) = \bigcup_{l = 1}^{\infty} \frac{1}{l} S^{(1)}(\CN^{\otimes l})
\label{esen}
\eeq
where $S^{(1)}(\CN)$ is the union, over all $\sigma^{X A B}$
arising from the channel $\CN$, of the $(r, R)$ pairs obeying
\begin{eqnarray}
0 & \leq & r \,\,\,\,\, \leq \,\,\,\, I(X; B)\nonumber \\
0 & \leq & R  \,\,\,\, \leq  \,\,\,\, I(A\,\rangle B X). 
\label{regija} 
\end{eqnarray}
Furthermore, in computing $S^{(1)}(\CN)$ one only needs to consider
random variables $X$ defined on some set $\CX$ of  cardinality
$|\CX| \leq (\dim \CH_{A'})^2 + 2$.
\end{thmc}

\begin{figure}
\centerline{ {\scalebox{.53}{\includegraphics{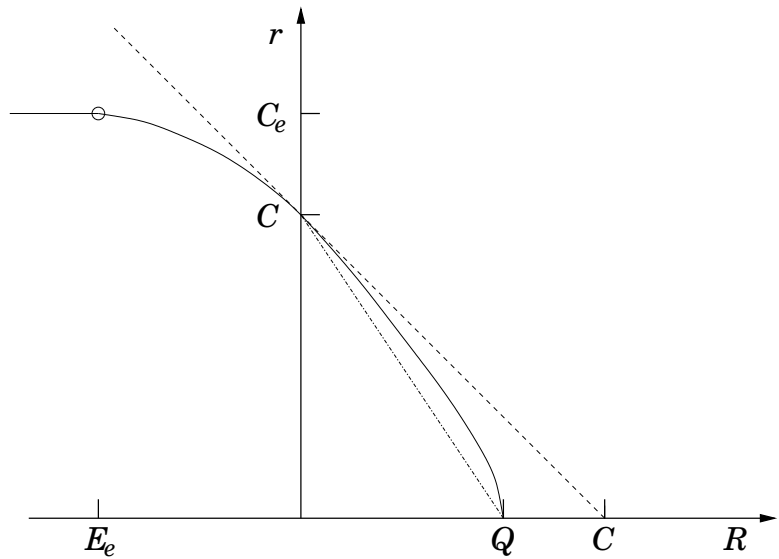}}}}

\caption{A generic trade-off curve for the simultaneous 
(classical, quantum) capacity region (solid line). 
The dashed-dotted line represents the time-sharing inner bound. 
The dashed line is the outer
bound which follows from the observation that the transmitted quantum 
subspace may always be used to encode classical information at 1 bit/qubit.
The continuation  to the negative $R$ axis (see text) 
is shown for scenario II (solid) and scenario I (dashed).
} 
\end{figure}
\vspace{2mm}

\begin{figure}
\centerline{ {\scalebox{.5}{\includegraphics{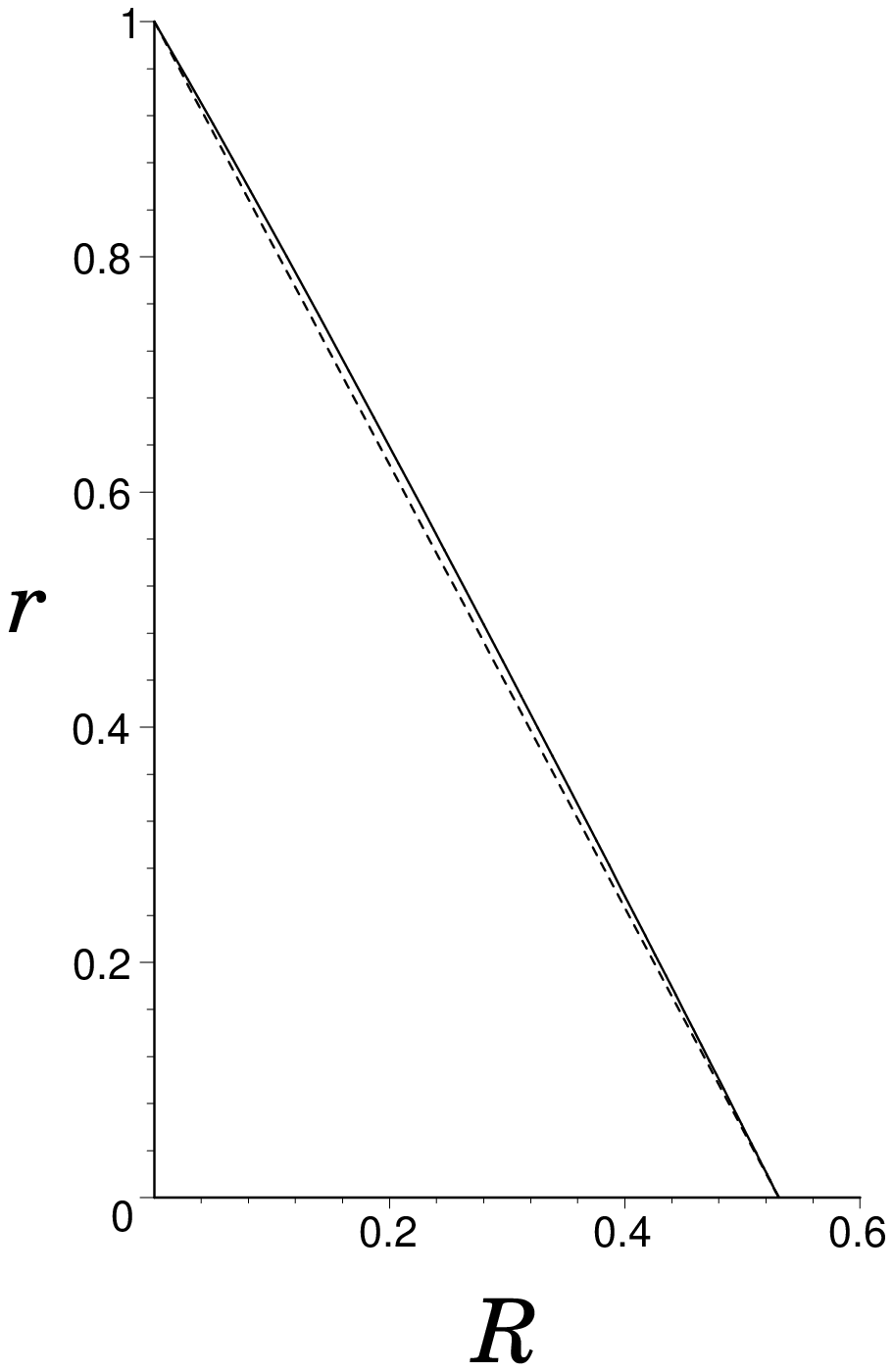}}}
 {\scalebox{.5}{\includegraphics{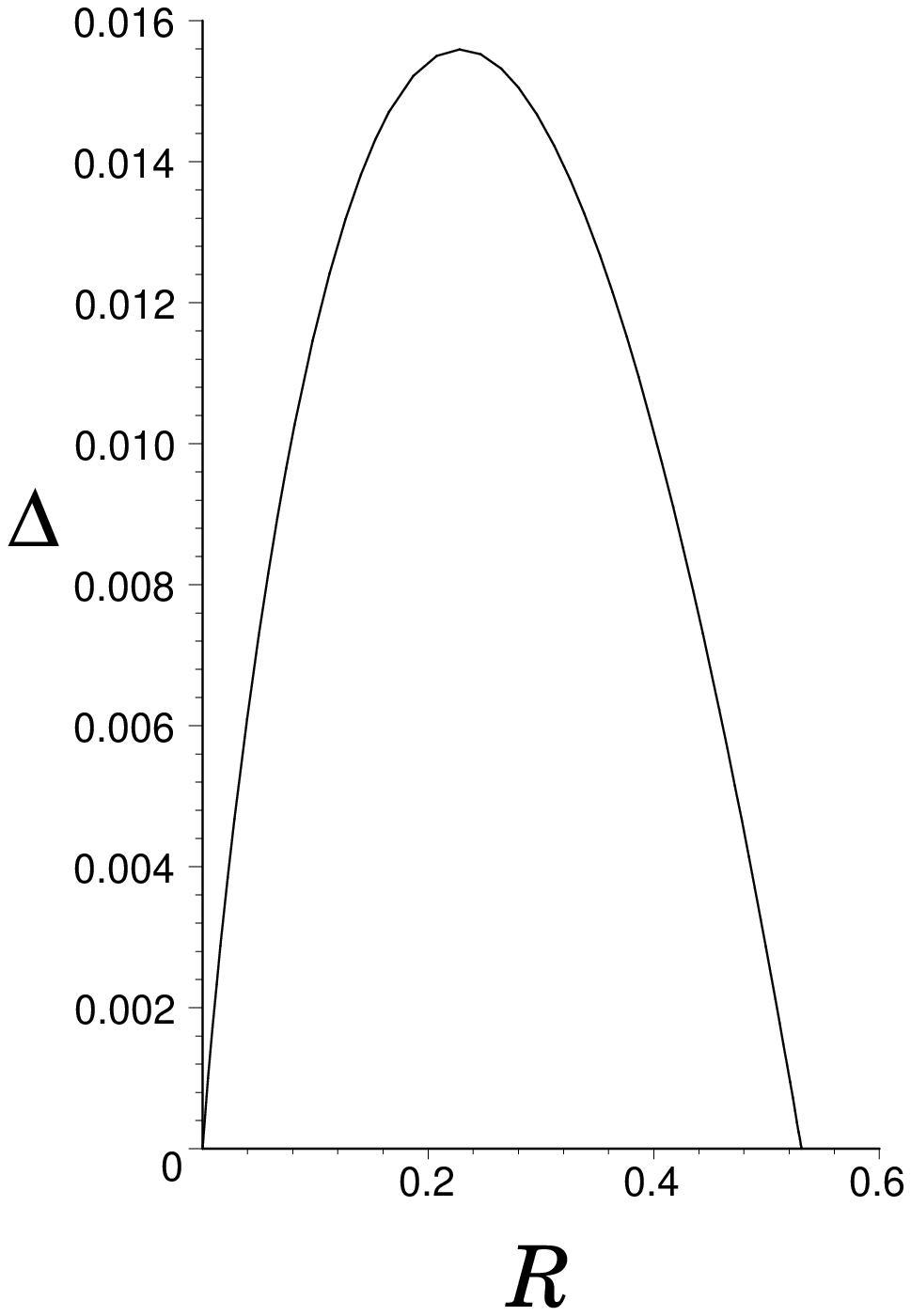}}}}

\caption{The trade-off curve for the dephasing qubit
channel with dephasing parameter 0.2 (i.e., the channel obtained
by applying the identity operator with probability 0.9 and $\sigma_z$ 
with probability 0.1).
In the left-hand plot, the trade-off curve is plotted with a
solid line and the
time-sharing bound with a dashed line. The right-hand plot
gives the difference between the optimal strategy and
time-sharing.
} 
\end{figure}
\vspace{2mm}

Since the three scenarios are equivalent we shall speak of a single capacity 
region. The generic shape of the capacity 
region is shown in figure 1. We shall informally 
refer to the outer boundary of the capacity region
in the $(r>0, R>0)$ quadrant as the ``trade-off curve''.
 In scenarios Ib and II,
for any $0 < \lambda < 1$,
combining a $(\lambda n, \epsilon)$ code of rate pair $(r_1, R_1)$ 
with a $((1 - \lambda) n, \epsilon)$ code of rate pair $(r_2, R_2)$ 
one obtains  an $(n, 2 \epsilon)$ code of rate pair 
$(\lambda r_1 + (1 - \lambda) r_2, \lambda R_1 + (1 - \lambda) R_2)$. 
This construction  is known as  \emph{time-sharing} and 
implies the concavity of the capacity region. 
In fact, the ``single-letter'' region $S^{(1)}(\CN)$ is already
concave for all channels $\CN$ (see appendix A).
The points $(C(\CN), 0)$ and $(0, Q(\CN))$ represent
the classical and quantum capacities, respectively. By time-sharing
one may achieve the line segment  interpolating between the two, 
giving an inner bound on the capacity region. An outer bound given
by the line segment connecting   $(C(\CN), 0)$ and $(0, C(\CN))$ is obtained
by observing that, in scenario Ia, the transmitted quantum subspace may 
always be used to encode classical information at 1 bit/qubit.

Our theorem is, alas, difficult to use in practice due to the 
$l \rightarrow \infty$ limit.
Two simple examples in which this limit is not 
required are the noiseless qubit channel and the erasure channel, for 
which both $C(\CN)$ and $Q(\CN)$ were previously known \cite{bds}. 
In both cases the boring time-sharing strategy turns out to be optimal. 
This is particularly trivial to see for the noiseless
channel: since $Q(\CN) = C(\CN)$, the inner and outer bound coincide.

A more interesting case is that of a dephasing channel, for which
the large $l$ limit is also not required (we prove this in appendix B),
yet the resulting trade-off curve is strictly concave. 
The $S(\CN)$ region for the dephasing qubit channel with dephasing 
parameter $0.2$ is shown in figure 2.  

For the depolarizing channel, another popular example, the 
$l \rightarrow \infty$ limit is known to be needed when the
depolarizing parameter is close to $p=0.189$, 
the value making $Q(\CN) = 0$ \cite{dss}. 
One can, however, make an interesting observation about 
the behavior of the trade-off curve near $R =  Q(\CN)$. Although the channel 
itself is
invariant under unitary transformations, the $\rho$ that maximizes the
coherent information $I_c(\rho, \CN)$ breaks this symmetry; indeed there 
is a whole family of density operators attaining $Q(\CN)$. One can thus construct an 
ensemble with $R =  Q(\CN)$ and $r>0$, so the trade-off curve is parallel to
the $r$ axis in a finite region around $r = 0$.
For the depolarizing channel with different $p$, we have calculated the
the trade-off curve assuming $l=1$ and found some interesting 
behavior.  For $p$ small ($p < 0.04$ or so) it is possible to do better 
than the time-sharing strategy, whereas for larger $p$ ($p > 0.05$), the 
time-sharing strategy is optimal, assuming $l=1$.  For these values of
$p$, it is not known whether taking large $l$ is advantageous for 
$Q(\CN)$.  

\vspace{2mm}

There is an intriguing connection between our capacity region and
the findings of Shor \cite{limit} concerning the classical capacity
of a quantum channel with limited entanglement assistance. 
The latter may be
thought of as extending  scenario II  to the 
negative $R$ axis, since entanglement is consumed rather than generated
\cite{aram}. The result for the $R \leq 0$ region parallels that from
theorem 1, replacing (\ref{regija}) by  
\begin{eqnarray} 
0 & \leq & r \,\,\,\,\, \leq \,\,\,\, I(X; B) + I(A;B|X)
\nonumber \\
& & R  \,\,\,\, \leq  \,\,\,\, -H(A|X) = I(A\,\rangle B X) - I(A;B|X). 
\label{legija} 
\end{eqnarray}
The two expressions on the right hand side
have the same sum as in equation (\ref{regija}).
There is a simple bijection between the two regions:
If $(r, R)$ is a point in $R \geq 0$ corresponding to
the state $\sigma^{X A B}$, then $(r + I(A;B|X), R - I(A;B|X))$ is
a point in the $R \leq 0$ region, and vice versa.
Imagine that 1 ebit of entanglement were a stronger resource that 
1 bit of communication, in the
sense that the latter could be produced form the former. 
Then the $R \leq 0$ region would be trivially achievable by the 
achievability of the $R \geq 0$ region. The opposite would hold were 1 bit 
stronger than 1 ebit. However, it is well known that bits and ebits are 
incomparable resources. 
The correspondence between the two regions may be interpreted as
providing a limited sense in which bits and ebits may be thought
of as equally strong. 
\par
One may play the same game in the context of scenario I (a or b),
with a somewhat less interesting outcome.
Here a negative rate $R$ is interpreted as assistance by a 
noiseless quantum channel. It is known \cite{swadd}
that the classical capacity of a noiseless channel combined with
a noisy one is just the sum of the individual capacities. Hence
the scenario I continuation of our trade-off curve simply 
follows the linear outer bound into the $R < 0$ region (see figure 1).

\vspace{2mm}

\section{Proof of theorem 1}

The following lemma from \cite{bkn} is needed to relate
scenarios Ia and Ib.

\begin{lemma1}
\label{napoli}
\begin{enumerate}
\item If 
$$
\min_{\ket{\varphi} \in \CH} F({\varphi},
 (\CD \circ \CN^{\otimes n} \circ \CE) ({\varphi})) 
\geq 1 -  {\frac{2}{3}} \epsilon$$
then
\beq 
F({\Phi}, [\1 \otimes (\CD \circ \CN^{\otimes n} \circ \CE)]
({\Phi})) \geq 1 - \epsilon.
\label{gurk}
\eeq
\item Conversely, if \emph{(\ref{gurk})} holds then
$$
\min_{\ket{\varphi} \in \CH'} F({\varphi},
 (\CD \circ \CN^{\otimes n} \circ \CE) ({\varphi})) 
\geq 1 -  2 {\epsilon},$$
where $\CH'$ is a subspace of $\CH$ satisfying
\beq
\dim \CH' \geq \frac{1}{2} \,  \dim \CH.
\label{mahu}
\eeq
\end{enumerate}
\end{lemma1}

\vspace{1mm}

Observe that $S_{\rm{Ia}}(\CN) = S_{\rm{Ib}}(\CN) \subseteq 
 S_{\rm{II}}(\CN)$.
The equality follows from both parts of lemma \ref{napoli}. 
The inclusion is obvious since one can always 
generate entanglement by transmitting half of the maximally entangled state
$\ket{\Phi}$. Therefore, to prove theorem 1 it suffices to show
that the region (\ref{esen}) is contained in $S_{\rm{Ib}}(\CN)$ (the 
``direct
coding theorem'') and contains $S_{\rm{II}}(\CN)$ (the ``converse''). 
\par
To prove the converse  we need the following simple lemma \cite{key}.
\begin{lemma2}
\label{bari}
For two bipartite states $\rho^{A B}$ and $\sigma^{A B}$ 
of a quantum system $AB$ of dimension $d$ with fidelity 
$f = F(\rho^{A B}, \sigma^{A B})$,
$$
|I(A \,\rangle B)_\rho - I(A \,\rangle B)_\sigma|
\leq \frac{2}{e} + 4 \log d \,  \sqrt{1 - f}.
$$
\end{lemma2}

\vspace{1mm}

\noindent{\bf Proof of theorem 1 (converse for scenario II)} \,\,\, 
Define the classical-quantum state $\omega^{M A B^n}$ to be
the result of sending the  $A'^n$ part of 
$$
\frac{1}{\mu} \sum_m 
\ket{m} \bra{m}^M \otimes \Upsilon_m^{A A'^n}
$$
through the channel $\CN^{\otimes n}$.
We shall prove that, for any $\delta, \epsilon > 0$ and all
sufficiently large $n$, if an $(n, \epsilon)$ code has a rate
pair $(r, R)$ then 
\begin{eqnarray}
r - \delta & \leq &  \frac{1}{n} I(M; B^n)_\omega, \label{ek}   \\
R - \delta & \leq & \frac{1}{n} I(A \,\rangle B^n M)_\omega. 
\label{do} 
\end{eqnarray}
Evidently, it suffices to prove this for $\delta \leq 1$, 
$\epsilon \leq 
[\frac {\delta}{16 \log \dim \CH_{A'}}]^2$ and $n \geq \frac{2}{ \delta}$.
Fano's inequality \cite{ct} says
$$
H(M|M') \leq 1 +  \Pr \{ M' \neq M \} n r.
$$
Equation (\ref{ek}) is a consequence of the following string of 
inequalities 
\begin{eqnarray*}
n r & = &  H(M) \\
& = & I(M;M')  + H(M|M') \\
& \leq & I(M; M') + 1 + n   \epsilon \log \dim \CH_{A'} \\
& \leq & I(M; B^n) + 1 + n \epsilon \log \dim  \CH_{A'},
\end{eqnarray*}
the last line by the Holevo bound \cite{holevo}. 
On the other hand, defining $\omega'^{M A B'}$ to be the
state $\omega^{M A B}$ after Bob's decoding $\CD$, 
\begin{eqnarray*}
I(A \,\rangle B^n M)_\omega
& \geq & I(A \,\rangle B' M)_{\omega'} \\
& \geq & I(A \,\rangle B')_{\omega'}  \\
& \geq & I(A \,\rangle B')_{\Phi}  -
 \frac{2}{e} - 8 n R \sqrt{\epsilon} \\
& \geq & n R - \frac{2}{e} - 8 n \log \dim \CH_{A'} \sqrt{\epsilon}, 
\end{eqnarray*}
from which the claim (\ref{do}) follows. 
The first inequality is the data processing inequality \cite{bns},
the second follows from the fact that conditioning cannot increase
quantum relative entropy \cite{nie&chuang} and
the third  is an application of lemma \ref{bari}.
It should be noted that we only used a weaker ``average'' version of
conditions 1. and 2$'$., namely
\begin{eqnarray*}
\lefteqn{  1. \, \Pr \{M'  \neq  M \} \leq  \epsilon, }
\,\,\,\,\,\,\,\,\,\,\,\,\,\,\,\,\,\,\,\,\,\,\,\,\,\,\,\,\,\,\,\,\,\,\,
\,\,\,\,\,\,\,\,\,\,\,\,\,\,\,\,\,\,\,\,\,\,\,\,\,\,\,\,\,\,\,\,\,\,\,
\,\,\,\,\,\,\,\,\,\,\,\,\,\,\,\,\,\,\,\,\,\,\,\,\,\,\,\,\,\,\,\,\,\,\,
\,\,\,\,\,\,\,\,\,\,\,\,\,\,\,\,\,\,\,\,\,\,\,\,\,\,\,\,\,\,\,\,\,\,\,
\,\,\,\,\,\,\,\,\,\,\,\,\,\,\,\,\,\,\,\,\,\,\,\,\,\,\,\,\,\,\,\,\,\,\,
\,\,\,\,\,\,\,\,\,\,\,\,\,\,\,\,\,\,\,\,\,\,\,\,\,\,\,\,\,\,\,\,\,\,\,
 \label{jen2} \\ 
\lefteqn{2'.  \, F ({\Phi}, \frac{1}{m} \sum_m\Omega_m ) \geq 1 - 
\epsilon. } 
\,\,\,\,\,\,\,\,\,\,\,\,\,\,\,\,\,\,\,\,\,\,\,\,\,\,\,\,\,\,\,\,\,\,\,
\,\,\,\,\,\,\,\,\,\,\,\,\,\,\,\,\,\,\,\,\,\,\,\,\,\,\,\,\,\,\,\,\,\,\,
\,\,\,\,\,\,\,\,\,\,\,\,\,\,\,\,\,\,\,\,\,\,\,\,\,\,\,\,\,\,\,\,\,\,\,
\,\,\,\,\,\,\,\,\,\,\,\,\,\,\,\,\,\,\,\,\,\,\,\,\,\,\,\,\,\,\,\,\,\,\,
\,\,\,\,\,\,\,\,\,\,\,\,\,\,\,\,\,\,\,\,\,\,\,\,\,\,\,\,\,\,\,\,\,\,\,
\,\,\,\,\,\,\,\,\,\,\,\,\,\,\,\,\,\,\,\,\,\,\,\,\,\,\,\,\,\,\,\,\,\,\,
\label{uvjet2}
\end{eqnarray*}
The bound on the cardinality of $\CX$ is proven in appendix C.
\qed

\vspace{1mm}

We henceforth restrict attention to scenario Ib.
In proving the  direct coding theorem, 
we shall combine purely quantum  and 
purely classical codes. 
A \emph{quantum code} is a
special case of a (classical, quantum) code defined
earlier, for which $\mu = 1$ ($r = 0$).
Quantum codes are characterized by a
pair of encoding and decoding maps $(\CE, \CD)$.
Define the \emph{quantum code density operator} \cite{key}
as $\CE(\pi)$, where $\pi = \frac{1}{ \kappa} \1^{A''}$.

Often in coding theory is it useful to consider \emph{random codes}.
Alice and Bob have access to an auxiliary resource: a common source of randomness
described  by some probability distribution $(P_\alpha)$.
A random quantum code is an ordered set of encoding-decoding pairs 
$((\CE^\alpha, \CD^\alpha))_\alpha$, indexed by $\alpha$.
With probability $P_\alpha$, Alice and Bob choose to
employ the deterministic code $(\CE^\alpha, \CD^\alpha)$.
The average code density operator for the random quantum code is given by 
$\sum_\alpha  P_\alpha  \CE^\alpha(\pi)$.
Given a density operator $\rho \in \CH_{A'}$, we say that an 
$(n, \epsilon)$ random quantum code is 
``$\rho$-type'' if the average code density operator $\omega$ satisfies
\beq
\|\omega - \rho^{\otimes n} \|_1 \leq \epsilon. 
\label{satis}
\eeq
For an ensemble of density operators $E = \{p_x, \rho_x \}$ 
defined on $\CH_{A'}$
and sequence $x^n = x_1 x_2 \dots x_n$ denote
$\rho_{x^n} = \bigotimes_{i=1}^{n} \rho_{x_i}$. 
We say that an $(n, \epsilon)$ random quantum code is ``$(E, x^n)$-type''
if the average code density operator $\omega_{x^n}$ satisfies
$$
\|\omega_{x^n} - \rho_{x^n} \|_1 \leq \epsilon. 
$$

The following proposition is a refinement of the 
quantum channel coding theorem, and was 
proved in Appendix D of \cite{key}.
A perhaps more accessible outline of the proof may 
be found in \cite{dw:prl}. 

\begin{prop1}
\label{pravi} 
For any $\epsilon, \delta > 0$ and all sufficiently large $n$,
there exists a random $\rho$-type  $(n, \epsilon)$ 
quantum code for the channel $\CN$ of rate $R = I_c(\rho, \CN) - \delta$.
\end{prop1}

Recall the notion of $\delta$-typical sequences $\CT^n_{p, \delta}$

$$\CT^n_{p,\delta}=\left\{x^n:\forall x\ |N(x|x^n)-n p_x|\leq
                               {\delta} n \right\},$$
where $N(x|x^n)$ counts the number of occurrences of $x$ in $x^n$.
When the distribution $p$ is associated with some random variable
$X$ the alternative notation $\CT^n_{X, \delta}$ may be used.  
Proposition (\ref{pravi}) extends to:

\begin{prop2} 
\label{pro2}
For any $\epsilon, \delta > 0$ and all
sufficiently large $n$, for any typical sequence 
$x^n \in  \CT^n_{p, \delta}$ there exists a 
random $(E, x^n)$-type $(n, \epsilon)$ 
quantum code for the channel $\CN$
of rate $R = \sum_x p_x I_c(\rho_x, \CN) - c \delta$, for some
constant $c$.
\end{prop2}


\vspace{2mm}

{\bf Proof} \,\,\, By  proposition \ref{pravi}, 
for sufficiently large $n$, for all $x$
there exists an  $(n[p_x - \delta], \epsilon)$ code
of rate $R_x = I_c(\rho_x, \CN) - \delta$,
with average density operator $\omega_x$ satisfying 
$$
\|\omega_x - \rho_x^{\otimes n[p_x - \delta]} \|_1 \leq \epsilon. 
$$
By ``pasting'' $|\CX|$ such codes together 
(one for each $x$) an 
$(n - |\CX| \delta, |\CX|\epsilon)$ code is produced with average code density
operator $\omega = \bigotimes_x \omega_x$.
Applying the triangle inequality multiple times,
\beq
\|\omega  - 
\bigotimes_x \rho_x^{\otimes n[p_x - \delta] } \|_1 \leq |\CX| \epsilon.
\eeq
Given $x^n \in  \CT^n_{X, \delta}$, abbreviate $n_x = N(x|x^n)$ and
$\Delta n_x = n_x -  n[p_x - \delta]$.
Now transform the above code into the ``padded'' $(n,  |\CX|\epsilon)$ 
quantum code 
obtained by inserting $\rho_x^{\otimes \Delta n_x}$ after each $\omega_x$;
its average density operator $\omega'$  obeys
\beq
\|\omega'  - \bigotimes_x \rho_x^{\otimes  n_x} \|_1 \leq |\CX| \epsilon.
\label{bond}
\eeq
The new rate $R$ is bounded by
$$
R  =  \sum_x R_x [p_x - \delta] \geq \sum_x p_x I_c(\rho_x, \CN) - 
\delta (1  + |\CX| \log \dim \CH_{A'}). 
$$
Finally, as $\bigotimes_x \rho_x^{\otimes  n_x}$ and $\rho_{x^n}$
are related by a permutation of the channel input Hilbert spaces 
and the channel $\CN^{\otimes n}$ is invariant
under such  permutations,
there  exists an $(n,  |\CX|\epsilon)$ code of the same rate $R$
and average code density operator $\omega_{x^n}$ such that
$$
\|\omega_{x^n}  - \rho_{x^n} \|_1 \leq |\CX| \epsilon.
$$ 
\qed

\vspace{2mm}

On the classical
side, we shall need the Holevo-Schumacher-Westmoreland (HSW) 
theorem \cite{h, sw}, or rather its ``typical codeword'' version
\cite{key}. Consider the restriction of $\sigma^{XAB}$ 
(\ref{szigma}) to $XB$:
$$
\sigma^{XB} = \sum_{x \in \CX} p_x \ket{x} \bra{x}^X \otimes 
\CN(\phi_x^{A'})^{B}.
$$
\begin{prop0} [HSW Theorem]
\label{hsw}
For any $\epsilon, \delta > 0$,
define $r = I(X; B)  - c' \delta$, for some constant $c'$,
and $\mu = 2^{nr}$.
For all sufficiently large $n$,
there exists a classical encoding map 
$f: [\mu] \rightarrow \CT^n_{X, \delta}$ and a decoding POVM
$\Lambda = (\Lambda_{m})_{m \in [\mu] }$, such that
$$
\tr \tau_m \Lambda_{m} \geq 1 - \epsilon, \,\,\,
 \forall m \in [\mu],
$$
where 
$$
\tau_m = \CN^{\otimes n} (\phi^{A'^n}_{f(m)})
$$
and $\phi^{A'^n}_{x^n} = \bigotimes_{i=1}^{n} \phi^{A'}_{x_i}$.
\end{prop0}

\vspace{2mm}
Proposition \ref{hsw} says that Bob may reliably distinguish 
among $\mu$ states 
 of the form 
$\CN^{\otimes n}(\phi^{A'^n}_{x^n})$, with $x^n \in \CT^n_{X, \delta}$. 
The idea behind the 
proof of the direct coding theorem 
is for Alice to use a different quantum code 
depending on the classical message to be sent.
Bob first decodes the classical message
(while causing almost no disturbance to the quantum system)
by taking advantage of the distinguishability of
the channel outputs for the different codes.
Furthermore, the same information tells him which quantum decoding 
to perform! Thus, the classical information has been ``piggy-backed''
on top of the quantum information.

\vspace{2mm}

\noindent{\bf Proof of Theorem 1 (coding for scenario Ib)} \,\,\,
Recall, in scenario Ib Alice is transmitting half of the 
maximally entangled state $\ket{\Phi}$ through the channel.
Define $\mu$, $f$, $\tau_m$ and $\Lambda$ as in 
proposition \ref{hsw}. For now we shall assume  Alice and Bob have
access to a common source of randomness with distribution $(P_\alpha)$.  
For each $m$ define a  $(\{p_x, \phi^{A'}_x \}, f(m))$-type 
$(n, \epsilon)$ random quantum code of rate 
$R = I(A\,\rangle B X) - c \delta$
by the encoding and decoding operators 
$((\CE_{m}^\alpha, {\CD'}_m^\alpha))_\alpha$.
By proposition \ref{pro2} and monotonicity of trace distance \cite{nie&chuang}
we have, for all $m$ and sufficiently large $n$,
$$
\|\sum_\alpha P_\alpha{\tau'}^\alpha_m - \tau_m \|_1 
\leq \epsilon, 
$$
where ${\tau'}^\alpha_m = (\CN^{\otimes n} \circ  \CE_m^\alpha) (\pi)$.
By proposition  \ref{hsw},
\beq
\sum_\alpha P_\alpha \tr {\tau'}^\alpha_m \Lambda_{m} \geq 1 - 2 \epsilon. 
\label{skopi}
\eeq
For a specific value of $\alpha$,
the encoding map for our (classical, quantum) code 
is given by $(\CE_m^{\alpha})_{m \in [\mu]}$. The
decoding instrument $\mathbf{D}^\alpha$ is given by 
$$
\CD_m^\alpha: \rho \mapsto {\CD'}_m^\alpha 
(\sqrt{\Lambda_{m}} \rho \sqrt{\Lambda_{m}}).
$$
As usual, $\CD^\alpha = \sum_m \CD_m^\alpha$ denotes the induced quantum decoding 
operation.
By (\ref{skopi}), for all $m$,
\beq
\sum_\alpha P_\alpha \, p_{{\rm err}}^{m, \alpha} \leq 2 \epsilon,
\label{pera1}
\eeq
where $p_{{\rm err}}^{m, \alpha} = 1 - \tr \CD^\alpha_m({\tau'}^\alpha_m) $.
 Defining an extension of ${\tau'}^\alpha_m$
$$
\xi^\alpha_m = [\1 \otimes (\CN^{\otimes n} \circ \CE^\alpha_m)]
 (\ket{\Phi} \bra{\Phi}),
$$
it follows from (\ref{skopi}) that
$$
\sum_\alpha P_\alpha \tr \xi^\alpha_m (\1 \otimes \Lambda_m) \geq 1 - 2 \epsilon.
$$
Invoking the gentle measurement lemma \cite{strong}
and the concavity of the square root function,
$$
\sum_\alpha P_\alpha \| (\1 \otimes \sqrt{\Lambda_m}) \xi_m (\1 \otimes \sqrt{\Lambda_m}) - 
\xi^\alpha_m \|_1 \leq 4 \sqrt{\epsilon},
$$ 
which by the monotonicity of trace distance \cite{nie&chuang} gives 
$$
\sum_\alpha P_\alpha \| (\1 \otimes \CD_m) (\xi^\alpha_m) - 
(\1 \otimes \CD'_m) (\xi^\alpha_m) \|_1 \leq 4 \sqrt{\epsilon}.
$$
On the other hand, 
$$
\| (\1 \otimes \CD) (\xi^\alpha_m) - 
(\1 \otimes \CD^\alpha_m) (\xi^\alpha_m) \|_1 \leq \sum_{m' \neq m}
 \|{\CD'}^\alpha_m(\xi^\alpha_m) \|_1 
\leq  2 {\epsilon}.
$$
Since, for all $m, \alpha$, 
$$
F((\1 \otimes {\CD'}^\alpha_m) (\xi^\alpha_m), \Phi) \geq 1 - \epsilon,
$$
putting everything together gives, for all $m$,
\beq
\sum_\alpha P_\alpha \, P_{{\rm err}}^{m, \alpha} \leq 3 \epsilon + 4 \sqrt{\epsilon}
\label{pera2}
\eeq
where 
$ P_{{\rm err}}^{m, \alpha} = 1 - F((\1 \otimes \CD^\alpha) (\xi^\alpha_m), \Phi) $.

\vspace{2mm}

At this point our code relies on Alice and Bob having access to 
the common random index $\alpha$.  
To prove the theorem it remains to ``derandomize'' the code, 
i.e. show that $p_{{\rm err}}^{m, \alpha}$ and 
$P_{{\rm err}}^{m, \alpha}$ are small for a particular value of $\alpha$, 
and for $m$ in a sufficiently large subset of $[\mu]$. 
By (\ref{pera1}) and (\ref{pera2}),
$$
\sum_\alpha P_\alpha 
\frac{1}{\mu} \sum_m (p_{{\rm err}}^{m,  \alpha} + P_{{\rm 
err}}^{m, \alpha}) \leq 5 \epsilon + 4 \sqrt{\epsilon}.
$$
There exists a particular $\alpha$ for which
$$
\frac{1}{\mu} \sum_m (p_{{\rm err}}^{m,  \alpha} + P_{{\rm 
err}}^{m, \alpha}) \leq 5 \epsilon + 4 \sqrt{\epsilon}.
$$
Fixing $\alpha$, expurgate the worst half of the codewords, 
i.e. those $m$ with the 
highest value of $p_{{\rm err}}^{m, \alpha} + P_{{\rm err}}^{m, \alpha}$. 
Now we have a code with both
$p_{{\rm err}}^{m, \alpha} $ and   $P_{{\rm err}}^{m, \alpha}$
bounded from above by $10 \epsilon + 8 \sqrt{\epsilon}$ 
for all remaining $m$, 
while the classical rate has only decreased by $\frac{1}{n}$.
This concludes the proof.
 \qed

\section{Remarks on related problems}

The first remark we make concerns replacing the classical--quantum dichotomy
with the cryptographically relevant public--private one.
In \cite{key} quantum codes were built
based on private information transmission ones. The purpose of the latter 
is for sending classical information about which the potential eavesdropper (to
which the ``environment'' of the channel is granted) cannot learn anything.
This should be contrasted with HSW codes which may be viewed
as transmitting public information. One may now consider the problem
of finding the simultaneous (public, private) capacity of $\CN$. 
The answer follows in a straightforward manner from the methods of 
$\cite{key}$ and those used in proving theorem 1. Viewing the channel
$\CN$ as being embedded in an isometry $U_\CN$ with an enlarged target
 Hilbert space,
$U_\CN: \CH_{A'} \rightarrow \CH_B \otimes \CH_E$ 
($\CH_E$ is now given to the eavesdropper), the simultaneous 
(public, private) capacity region is given by the following modification
of theorem 1:
\begin{itemize}
\item replace the state $\sigma^{XAB}$ by $\sigma^{XYB}$, obtained
by sending the $A'$ part of 
$$
 \sum_{xy} p_{xy} \ket{x}\bra{x}^X \otimes \ket{y}\bra{y}^Y 
\otimes \rho_{xy}^{A'}
$$
through the channel, 
\item replace $I(A \,\rangle B X)$ by 
$I(Y; B|X) -  I(Y; E|X)$.
\end{itemize}
The corresponding theorem for classical ``wire-tap''channels
was proven in \cite{ck2}.  
\vspace{2mm}

Secondly, one may conceive of a ``static'' analogue of the problem considered here,
where Alice and Bob share many copies of some (mixed) state $\rho^{AB}$
instead of being connected by a quantum channel.
In \cite{CR} the problem of generating common randomness (perfectly correlated bits)
from such a resource  using limited forward (Alice to Bob) classical communication 
was considered. There the ``distillable common randomness'' (DCR) was defined to 
be the maximum common randomness obtainable \emph{in excess} of the 
classical communication invested, and was advertised as an (asymmetric) 
measure of the classical correlations in $\rho^{AB}$.
In \cite{devetak:winter} the problem of one-way entanglement distillation
was solved, yielding a similarly asymmetric measure of \emph{quantum} correlations
in $\rho^{AB}$. 
The next step is to unify the two results in a 
trade-off between DCR and distillable entanglement, which
could now be argued to quantify the \emph{total} 
correlations in the state.
Based on the results of \cite{CR}, \cite{devetak:winter} and the present
paper we put forth the following conjecture: The simultaneously
distillable (classical, quantum) resources are given precisely by theorem 1,
where now the test states $\sigma^{XAB}$ are obtained by applying 
general instruments $\mathbf{D} = (\CD_x)_{x \in \CX}$ to the $A$ part of
$\rho^{AB}$, rather than arising from a channel. A sketch
of the proof is as follows. The coding strategy involves double blocking.
First use the protocol of \cite{CR} on a block of length $n$ to
establish a good approximation to $X^n$ on Bob's side using 
$\approx n H(B|X)$ bits of forward communication. This already
gives us the desired DCR rate of $I(X;B)$. Now that Bob's system
includes $X^n$ they may use further blocking to distill entanglement at a 
rate of $I(A\,\rangle BX)$ \cite{devetak:winter}. The classical
communication involved in this distillation has now turned into common 
randomness, effecting no net change in the DCR. The converse theorem
is left as an exercise.
A somewhat more ambitious goal
would be to include the classical communication cost in the trade-off,
giving a 3-dimensional region!

The final remark we make is that the ``piggy-backing'' idea
used in the proof of theorem 1
provides an alternative coding strategy to the one in \cite{limit} for
the classical capacity of $\CN$ with limited entanglement assistance,
thus establishing an additional connection between the two problems.
The original paper on the entanglement assisted capacity \cite{eac}
describes how to achieve the pair $(r, R) = (I(A;B)_\rho, -H(A)_\rho)$, for
some $\rho^{AB} = (\1^A \otimes \CN)(\phi^{A A'})$ arising from the channel.
Using a mixture of codes corresponding to different channel inputs
$\ket{\phi_x}^{A A'}$, one trivially
achieves $(r, R) = (I(A;B|X), - H(A|X))$ (with respect to $\omega^{XAB}$). 
As it turns out, Bob may use the distinguishability of the channel
outputs of different such code mixtures to send extra classical
information at a rate of $I(X;B)$. This gives the region (\ref{legija}).
A detailed version of this argument will appear in \cite{dhw}.
\vspace{2mm}

\section{Discussion} 
In conclusion, an information theoretical characterization
of the simultaneous (classical, quantum) capacity region has been derived.
The key idea was to use a different quantum code depending on the classical
information to be sent, thus ``piggy-backing'' the classical information 
on top of the quantum one. The formula derived requires optimization
over potentially arbitrarily many copies of the channel. We have
shown that for a generalized dephasing channel a single copy suffices.
We have also presented some ideas 
on cryptographic as well as static analogues of this problem.
 
We have already mentioned the open problem of   
including the classical communication cost in the trade-off for the static
analogue. 
Another interesting extension of our work, which in fact served as our
original motivation, is the following joint source-channel coding problem. 
In \cite{hjw} the task of quantum compression with classical
side information was considered. This is a ``visible'' source coding 
problem of a pure-state ensemble $E$. 
By storing partial information about the identity of the states
(classically) at a rate $C$ it is possible to reduce the 
quantum storage rate to some value $Q(C)$. The joint source-channel coding variant 
of this problem is: Given $E$ and a channel $\CN$, what
is the rate at which Alice can send the quantum part of the ensemble 
over the channel? One approach is to first separate the source into a 
classical and quantum part using the trade-off of \cite{hjw} and then send
them simultaneously through the channel using the trade-off of
theorem 1. This procedure is optimized over the ratio $\lambda$ of 
the classical and quantum rates  
which should coincide for the source and channel coding part.
There are, however ``well matched'' source-channel pairs for which such a strategy is 
known to be suboptimal.
The following example is due to Smolin \cite{john}.
The source is the equiprobable ``trine'' ensemble  
$(\ket{0}, \ket{\epsilon^+}, \ket{\epsilon^-} )$,
where $\ket{\epsilon^\pm} =  \frac{1}{2} \ket{0} \pm  \frac{\sqrt{3}}{2} \ket{1}$
and the channel $\CN: \CH_3 \rightarrow \CH_2$ has operation elements 
$\{ \ket{0}\bra{0} , \ket{\epsilon^+} \bra{1}, \ket{\epsilon^-} \bra{2} \}$.
The channel has no quantum capacity and a classical capacity of $1$.
Our strategy of separating the source and channel coding gives a source-channel
capacity of $1/\log 3$ transmitted copies of 
the ensemble per channel use. On the other hand, by simply feeding the identity
of the state to the channel one achieves a source-channel capacity
of $1$. Finding a solution for an arbitrary $(E, \CN)$ pair
remains an open question.

\noindent {\bf Acknowledgments} \,\,\, We thank Charles Bennett, 
Aram Harrow and John Smolin for useful discussions. 
ID is partially supported by the NSA under the 
ARO grant numbers DAAG55-98-C-0041 and DAAD19-01-1-06.

\vspace{2mm}

\appendix 

\section{Proof of concavity of $S^{(1)}(\CN)$}

Here we provide a proof that the region $S^{(1)}(\CN)$
defined by (\ref{regija}) is concave. Let $\sigma_0^{XAB}$ and 
$\sigma_1^{XAB}$ be two different states arising 
from the channel. For some $\lambda$ between $0$ and $1$,
consider the state
$$
\sigma^{UXAB} = \lambda \, \proj{0}^U \otimes \sigma_0^{XAB} +
(1 - \lambda) \, \proj{1}^U \otimes \sigma_1^{XAB}, 
$$
which also arises from the channel.
Then
\begin{eqnarray*}
\lambda \, I(X;B)_{\sigma_0} + (1 - \lambda)\, I(X;B)_{\sigma_1}
& \leq & I(UX;B)_\sigma \\
\lambda \, I(A \,\rangle BX)_{\sigma_0} + (1 - \lambda) \,
 I(A \,\rangle BX)_{\sigma_1}
& = & I(A \,\rangle BUX)_\sigma, 
\end{eqnarray*}
from which the claim follows.

\section{The capacity region for dephasing channels}
In this section we define the notion of 
\emph{degradable channels} and show that for such 
channels the quantum capacity is given by the single-letter formula
\beq
Q(\CN) = Q^{(1)}(\CN) := \max_{\rho^{AB}} I(A \, \rangle B),
\eeq
where the maximization is over all states $\rho^{AB}$
arising from the channel $\CN$.
For the special case of \emph{dephasing}
channels we shall prove that the entire trade-off 
curve can be single-letterized.

Recall that a channel $\CN: \CH_{A'} \rightarrow \CH_B$ can be defined by an 
isometric embedding $U_\CN: \CH_{A'} \rightarrow \CH_B \otimes \CH_E$, 
followed by a partial trace over the ``environment'' system $E$, 
so ${\cal N}(\rho) = \tr_{\!E} U_\CN(\rho)$ \cite{stinespring}. 
This further
induces the \emph{complementary channel} $\CN^c: \CH_{A'} \rightarrow \CH_E$
defined by ${\CN^c}(\rho) = \tr_{\!B} U_\CN(\rho)$.
We call a  channel $\CN$  \emph{degradable} when it may be degraded
to its complementary channel $\CN^c$, i.e. when there exists a map 
${\cal T}: \CH_{B} \rightarrow \CH_E $ so that $\CN^c = \CT \circ \CN$.

To see that $Q(\CN) = Q^{(1)}(\CN)$ for degradable channels,
note that Bob's output system $B$ may be mapped by a fixed isometry
onto a composite system $B' E'$ such that the
channels from $A'$ to $E'$ and to $E$ are the same. 
Thus, for any state arising from the channel,
\begin{eqnarray*}
I(A\rangle B) &=& H(B) - H(E) \\
&=& H(B'E') - H(E) \\
&=& H(B' E') - H(E') \\
&=& H(B' | E').
\end{eqnarray*}
We can then use the inequality \cite{nie&chuang}
\[
H(B'_1 B'_2| E'_1 E'_2) \leq H(B'_1|E'_1) + H(B'_2|E'_2)
\]
to prove that single-letter maximization already achieves $Q(\CN)$.

A subclass of degradable channels of particular interest 
are generalized dephasing channels. The latter are defined on some $d$-dimensional
Hilbert space with a preferred orthonormal basis $\{ \ket{i} \}$, 
such that all states belonging to this basis are transmitted without error, but 
pure superpositions of these basis states may become mixed.
This implies that if $\CN$ is a dephasing channel then its
isometric embedding $U_\CN$ obeys
$$
U_\CN \,\ket{i}^{A'} = \ket{i}^B \ket{\phi_i}^E, 
$$
where the $\ket{\phi_i}$ are generally not mutually orthogonal.
When the $\ket{\phi_i}$  are mutually orthogonal $\CN$
is the completely dephasing channel $\Delta_d$:
$$
\Delta_d (\rho) = \sum_{i = 1}^d \proj{i} \rho \proj{i}.
$$
It is clear from the above that any  dephasing channel 
$\CN$ obeys
\begin{itemize}
\item $\Delta_d \circ \CN = \CN \circ \Delta_d = \Delta_d$ 
\item $\CN^c \circ \Delta_d = \CN^c$ 
\end{itemize}


Every dephasing channel is degradable, since
$\CN$ may be degraded to $\Delta_d$ which may be further
degraded to $\CN^c$. In fact, the map $\CT$ can be taken to be $\CN^c$.
Therefore, $Q(\CN) = Q^{(1)}(\CN)$
In what follows, the special properties of dephasing channels will 
allow us to prove an even stronger statement: that the outer 
boundary of $S(\CN)$ may be expressed as a single-letter formula.

Consider some state
$\sigma^{X A B E}$ arising from the channel.
Bob may degrade his channel further by 
replacing  his system $B$ by $B'Y$, where $Y$ now
contains the completely dephased version of $B$ (this is why we label
it as a classical system).
Set $\lambda \geq 1$ and  define
$$
f_\lambda(\CN) = \max_{\sigma^{XYE}}
\left[ H(Y) + (\lambda - 1) \, H(Y|X) - \lambda \,H(E|X) \right ], 
$$
where the maximization is over all $\sigma^{XYE}$ arising from the channel 
($\sigma^{XYE}$ is implicit in the entropies).
We shall make use of  the following lemma.
\begin{lemma4}
For two general dephasing channels $\CN_1$ and $\CN_2$ 
$$
f_\lambda (\CN_1 \otimes \CN_2) = 
 f_\lambda (\CN_1) + f_\lambda (\CN_2)  
$$
\end{lemma4}

\noindent{\bf Proof} \,\,\,
The ``$\geq$'' direction follows from the fact that
the input ensemble for $\CN_1 \otimes \CN_2$ may be
chosen to be a tensor product of the ones 
maximizing $f_\lambda (\CN_1)$ and $f_\lambda (\CN_2)$.
To show the opposite inequality, in what follows
 let us refer to the state  $\sigma^{XY_1Y_2E_1E_2}$ 
that maximizes $f_\lambda (\CN_1 \otimes \CN_2)$.
Observe that 
\begin{eqnarray*}
H(Y_1 Y_2) & =  & H(Y_1) + H(Y_2|Y_1) \\
H(Y_1 Y_2|X) & =  & H(Y_1|X) + H(Y_2|Y_1 X) 
\end{eqnarray*}
and
\begin{eqnarray*}
H(E_1 E_2|X) & =  & H(E_1|X) + H(E_2|E_1 X) \\
 & \leq & H(E_1|X) + H(E_2| Y_1 X),
\end{eqnarray*}
the latter since $E_1$ contains a degraded version of $Y_1$
for all values of $X$.
Hence 
\begin{eqnarray*}
f_\lambda(\CN_1 \otimes \CN_2) & = & 
 H(Y_1 Y_2) + (\lambda - 1) \, H(Y_1Y_2|X) - \lambda \,  H(E_1E_2|X) \\
& \leq &   H(Y_1) + (\lambda - 1) \, H(Y_1|X) - \lambda \, H(E_1|X) \\
& + & H(Y_2|Y_1) + (\lambda - 1) \, H(Y_2|X Y_1) - \lambda \, H(E_2|X Y_1) \\
&  \leq &  f_\lambda (\CN_1) + f_\lambda (\CN_2).  
\end{eqnarray*}
\qed 

\vspace{3mm}

We shall use Lagrange multipliers to calculate $S(\CN)$. 
By theorem 1, the quantity to be maximized is
$$
I(X; B) + \lambda \, I(A \,\rangle BX),
$$ 
over all states $\sigma$ that arise from $\CN^{\otimes n}$.
Operationally it is clear that we should restrict attention to $\lambda \geq 1$,
since $-\lambda$ is the
the slope of the boundary of $S(\CN)$ and a qubit channel
may always be used to send classical bits at a unit rate.
For any such state we have
\begin{eqnarray*}
I(X; B) + \lambda I(A \,\rangle BX) & =  &  
H(B) + (\lambda - 1) H(B|X) - \lambda H(E|X) \\
& \leq  &  H(Y) + (\lambda - 1) H(Y|X) - \lambda H(E|X) \\
& \leq &  f_\lambda (\CN^{\otimes n}) \\ 
& \leq &  n f_\lambda(\CN).
\end{eqnarray*} 
The first inequality follows from the fact that complete dephasing
increases entropy, and is saturated by completely dephasing the 
\emph{input} to $\CN^{\otimes n}$ (recall that $\CN$ commutes with $\Delta_d$). 
The third inequality is by lemma 6.
Thus, for dephasing channels,
$S(\CN) = S^{(1)}(\CN)$, which makes the optimization problem
tractable.

We now turn to the particular case of the qubit $p$-dephasing channel
$$
\CN = (1 - p) \, \1_2 + p \, \Delta_2.
$$
It is easily checked that the outer boundary of of $S(\CN)$ is achieved by the 
$\mu$--parametrized family of ensembles, 
$\mu \in [0, 1/2]$, consisting of $\diag(\mu, 1 - \mu)$ and $\diag(1 - \mu, \mu)$ 
chosen with equal probabilities. 
The trade-off curve is given by
$$
(r, R) = \left( 1 - h_2(\mu),  h_2(\mu) - h_2 \left(1/2 + 1/2
\sqrt{1 - 16 p (1 - p) \mu (1 - \mu)} \right) \right),  
$$
where $h_2(\mu) = - \mu \log_2 \mu - (1 - \mu) \log_2 (1 - \mu)$ 
is the binary entropy function. Figure 2 shows this curve for $p = 0.2$.

\section{Proof of the cardinality bound}

Here we justify the condition on the cardinality of $\CX$ in 
the statement of theorem 1. 
Caratheodory's theorem states that
in a $t$-dimensional Euclidean space, each point of a connected compact
set $\CK$ can be represented as a convex combination of at most $t+1$
points in $\CK$. Let $\CF(\CH)$ be the family of all density
operators on the Hilbert space $\CH_{A'}$ of dimension $d$. Let $\CK$ be the 
image of $\CF(\CH)$ under some continuous mapping $f$ defined
by $f(\rho) = (f_1(\rho), \dots, f_t(\rho))$. As $\CF(\CH)$ 
is connected and compact, so is $\CK$. Then for any probability
measure $\mu$ on the algebra of density operators of $\CH_{A'}$, 
Caratheodory's theorem implies 
the existence of some finite ensemble $\{ p_x, \rho_x: x \in \CX \}$, $|\CX| = t + 1$, 
such that
$$
\int_{\CF(\CH)}  \mu(d \rho) f_j(\rho)  = 
\sum_{x \in \CX} p_x f_j(\rho_x), \,\,\,\,\, \forall j \in [t].
$$
Turning to our problem, the quantities $I(X;B)$ and
$I(A \, \rangle BX)$ depend solely on the ensemble
$E = \{ p_x,  \rho_x \}$, where $\rho_x := \phi^{A'}_x$, and
the channel $\CN$. Moreover, they only depend on the 
vector $\sum_x p_x f(\rho_x)$, where the vector valued function $f$
is defined so that
$f_1, \dots , f_{d^2-1}$ are the $d^2 - 1$ degrees of freedom of 
$\rho$ (linear in $\rho$),
$f_{d^2}(\rho) = H(\CN(\rho))$ and  
$f_{d^2+1}(\rho) = I_c(\rho, \CN)$. 
Suppose that a particular point in $S^{(1)}(\CN)$ corresponds to
some ensemble $E' = \{ \mu(d \rho),  \rho \}$.
The above implies that the same point is achievable by a
finite ensemble with at most $d^2 + 2$ elements.

\end{document}